\documentstyle[12pt, aaspp4, graphics]{article}

\begin{document}

\title{The Late Time Light Curve of SN 1998bw Associated with GRB980425}
\author{Eric H. McKenzie \altaffilmark{1}, and Bradley E.
Schaefer\altaffilmark{2}
\altaffiltext{1}{eric.mckenzie@yale.edu}
\altaffiltext{2}{schaefer@grb2.physics.yale.edu}
}
\affil{Department of Physics, Yale University\\
260 Whitney Avenue, JWG 463, New Haven, CT 06520-8120}

\baselineskip 24pt

\begin{abstract}

We report 139 photometric observations through the B, V, and I filters of
the supernova SN 1998bw, an object which is associated with the Gamma-Ray 
Burst GRB 980425. Detailed light curves of this unique supernova can be
compared to theoretical models, so we report here our light curve for 123
days between 27 June 1998 and 28 October 1998. The light curve of SN 1988bw
is consistent with those of the Type Ic class. We find that the
magnitude-versus-time relation for this supernova is linear to within 0.05
mags in all colors over
the entire duration of our study. Our measured uniform decline rates are 
$0.0141 \pm 0.0002$, $0.0184 \pm 0.0003$, and $0.0181 \pm 0.0003$
magnitudes per day in the B, V, and I bands. The linear decline and the
rate of that decline suggest that the late time light curve is powered by 
the radioactive decay of cobalt with some leakage of the gamma rays.
\end{abstract}

\keywords{supernovae: general; supernovae: individual 1998bw; 
gamma-rays: bursts}

\clearpage
\section{Introduction}

In their search for an optical counterpart for the gamma ray burst
GRB980425, Galama et al. (1998a) detected SN 1998bw in the galactic 
arm of ESO 184-G82(EOP 184-82), which Tinney et al. (1998) determined
to have a red shift of $0.0085 \pm 0.0002$. The supernova's light 
curves rose sharply after the burst, and its spatial coordinates were
well within the burst's error box, strongly suggesting a connection 
between the two events. The probability of their independence was
estimated by Galama et al. (1998a) at $1.1 \times 10^{-4}$. However,
BeppoSAX also detected a fading x-ray
source (generally thought to be the hallmark of the burst counterpart)
at a position inconsistent with SN 1998bw (Pian et al. 1998, Piro et
al. 1998, Pian et al. 1999), so the relationship between SN 1998bw and 
GRB980425 is unclear. 
Further observations showed that SN 1998bw is
positionally coincident with a second BeppoSAX x-ray source which has 
faded by a factor of two in brightness from 26 April to 10 November 1999,
which is consistent with x-ray emission from a supernova plus the galaxy
(Pian et al. 1999).

SN 1998bw has peculiar and unique properties other than a possible
association with a Gamma-Ray Burst. Its spectrum is unique (although
two Type Ic supernovae have somewhat similar spectra; see Iwamoto 
et al. 1998) and displays ejection velocities measured from the blue wings
of the Ca II line as high as $60, 000 km \cdot s^{-1}$ (Kulkarni et al.
1998). Its emissions at radio wavelengths increased much more quickly 
than other supernovae, and it is also the most luminous supernova to date at
radio wavelengths (Kulkarni et al. 1998). These coincidences of unusual
properties greatly strengthen the connections between GRB 980425 and SN
1998bw. In general, a concensus has emerged that the burst is related to
the supernova, and this has inspired much research detailing connections
between the two phenomena.

Due to the unique and pivotal nature of SN 1998bw, it is imperative that
the light curve be tracked in a wide range of optical bands as long as
possible. Galama et al. (1998a) tracked the U, B, V, R, and I light curves
for 58 days after the burst, and these showed a typical peak as generally
seen for supernovae of many types. In the interests of recording
as much data as possible for such an unprecedented event, we followed up
on their results with further observations in the B, V, and I filters.

\section{Observations}

The data were obtained using the Yale 1-m telescope, at the Cerro Tololo
Inter-American Observatory in Chile, from 27 June through 28 October 1998.
Our series of observations commenced as soon as the refurbishing of the
Yale 1-m telescope had been compeleted and the CCD camera installed. The
images' pixel size was 0.30'', with a field of view of $10.2' \times
10.2'$. Our exposure times were always 300 seconds per image, using B,
V, and I filters. Our typical seeing had a FWHM of $1.2''$. We obtained 
139 measures of the brightness of SN 1998bw.

The images were first processed with the normal procedure for overscan
correction, bias subtraction and flat fielding. Our photometric analyses
were made with IRAF's program ``APPHOT''. We used apertures of only three
pixel radius to minimize interference from the parent galaxy, which was 
a sufficient size because of the high signal-to-noise ratio. Our
background annuli were constructed with inner and outer radii of thirty
and forty pixels centered on the star, with the sky background taken as
the mode within this annulus. The deduced sky background for the supernova
is close to that deduced for isolated stars. 

For each image, observed magnitudes were recorded for SN 1998bw and for
five comparison stars. We used the updated magnitudes provided
by Galama et al. (1998b), choosing numbers 4, 6, 7, 9, and 10 because they
were relatively bright. Comparison star \#2 was excluded because its data
was just eratic enough to arouse suspicion that it might be a small
amplitude variable star, although is was not definitively identified 
as such.

We were able to estimate our magnitude uncertainties by comparing the 
standard stars with each other in a variety of images. These errors 
were determined to be substantially dependent on the apparent magnitudes 
of the stars due to the normal Poisson variations, which is why we used
the brightest ones for our data analysis. Monitoring the differences
between the standard stars in each image also enables us to catch 
photometric problems with the standard stars (due to cosmic rays, bad
columns, etc.). A few nights had large uncertainty due to clouds or bad
seeing. Our results are that in general we have systematic uncertainty of
0.02 mag added in quadrature with the statistical errors reported by 
IRAF. For the supernova, the statistical errors are generally
substantially smaller than our systematic errors in the early portions of
our light curve. The comparison stars are fainter than the supernova, yet
our use of the average of five stars as our `standard' improves the
accuracy of this `standard' to $\sim 0.01$ mag. In all, the uncertainties
in our supernova magnitudes typically range from 0.02 to 0.04 mag.

SN 1998bw appears in the spiral arm of its host galaxy, so we must
consider the effects of the galaxy light in our photometry. Fortunately,
the supernova was quite bright during the entire duration of our study and 
the contribution of light from the spiral arm is minimal. To be quantitative, 
we have measured the surface brightness of the center of the spiral arm 
on both sides of the supernova and compared this with the total brightness
within our photometric aperture centered on the supernova. Images of the 
galaxy from before the supernova show the brightness along the spiral arm
to be uniform along the position of the supernova, so we know that there
are no significant knots or stars at the supernova position. At the
beginning of our light curve, the contamination from galaxy light in 
our photometry aperture varied from $0.5 \%$ to $0.8 \%$ for the three
filters. So we have a systematic error which is smaller than our quoted
uncertainties that will make the supernova slightly fainter than
tabulated. Ideally, we should wait several years for the transient to 
fade to invisibility, then get further images with our same equipment and
subtract off the galaxy light; but in the meantime the systematic error
is known to be small. 

For each image, we compared the instrumental magnitude of the supernova to
the average instrumental magnitude of the five standard stars. This
difference was then applied to the average of the standard stars' actual
magnitudes taken from Galama et al. (1998b) to determine the actual magnitude
 of SN 1998bw. Our results are plotted in Fig. 1 (along with Galama's
earlier results) and tabulated in Table 1. 

We are impressed with the remarkable linearity of our portion of the light
curves. The best fit lines to our data are displayed in Figure 1, and we
see no significant systematic deviation from perfect lines at any time
or in any color. Our limits on systematic deviations are $<0.05$ for our entire
123 day observation time. In the B filter, the light is declining at 
$0.0141 \pm 0.0002$ magnitudes per day, which corresponds to a radioactive 
half-life 
of $53.4 \pm 0.8$ days. For V, these figures are $0.0194 \pm 0.0003$
magnitudes per day, which a corresponding radioactive half-life of $40.9
\pm 0.7$ days. For I, the figures are $0.0181 \pm 0.0003$ magnitudes per
day and a half-life of $41.6 \pm 0.7$ days.

The B light curve has a somewhat slower decay than in the V and I bands.
For an extinction of $A_{v} = 0.2$ (Galama et al. 1998), the
supernova's B-V was 0.82 mag at the beginning of our observation
period and around 0.3 mag toward the end. The extinction corrected V-I rose 
by a small amount, from 0.53 to 0.60 mag, during the same time period. 

 With our B, V, and I light curves, we can approximate the
bolometric light curve for radiation from the ultraviolet to the infrared. 
We have done this by first correcting for galactic extinction ($A_{v} = 0.20$),
converting our magnitudes into $f_{\nu}$, adopting a power law spectrum
from B
to V and from V to I, adopting a Rayleigh-Jeans spectrum for lower
frequencies than I, adopting a Wien spectrum for higher frequencies than
B, and integrating the spectrum. For JD2450996 (68 days after the burst)
we get a bolometric flux of $1.1 \times 10^{-11} erg \cdot cm^{-2}
\cdot s^{-1}$ or a  bolometric luminosity of $2.0 \times 10^{42} erg
\cdot s^{-1}$. For JD2451098  (170 days after the burst)
we get a bolometric flux of $2.1 \times 10^{-12} erg \cdot cm^{-2}
\cdot s^{-1}$ or a bolometric luminosity of $3.9 \times 10^{41} erg
\cdot s^{-1}$.  For the conversion 
to luminosity, we adopted a velocity of $2550 km \cdot s^{-1}$ and a 
Hubble Constant of $65 km \cdot s^{-1} \cdot Mpc^{-1}$, for a distance of 
39 Mpc.  These calculated luminosities have
significant uncertainties arising from the extinction ($\sim 10\%$), 
bolometric correction ($\sim 30\%$), and distance ($\sim 20\%$), so that 
overall errors perhaps as large as $\sim 50\%$ might be present. The 
effective half-life for this decline is 44 days.

\section{Comparison with Other Supernova}

SN1998bw has many unique and extreme properties, however at first
look, its light curve appears to be that of a normal supernova.  Is the
light curve unique?  We will compare our light curve with those of Type
Ia, Ib, Ic, and II supernovae in turn.
        
For the majority of Type Ia events, the B light curve fades by
1.1 magnitudes in the first 15 days afters peak (Hamuy 1996a), while
SN 1998bw has the same drop. The usual slope of the late time B light
curve is $0.01516 \pm 0.0024 mag/day$ for Type Ia events (Barbon et al.
1984) and is easily compatible with that of SN 1998bw. However, the
decline rates in V and I differ substantially between most of the Type Ia
events and SN 1998bw (0.0184 versus 0.024 and 0.0181 versus 0.041
magnitudes per day respectively) from 70-80 days after peak. A second
important difference is that almost all Type Ia events display a prominent 
bump in the I band light curves from 20-50 days after peak (Hamuy et al. 
1996a, Riess et al. 1999), while SN 1998bw does not show any sign of such 
a bump. A third difference is that SN 1998bw has a peak absolute magnitude 
of $-18.88 \pm 0.05$ (Galama et al. 1998a) whereas the majority of Type Ia
events have peak absolute magnitudes of $- 19.26$ (Hamuy et al. 1996b).
However, uncertainties in distance and extinction can perhaps be as large 
as a third of a magnitude, so this third difference may not be
significant. A fourth difference is that the extinction corrected color 
at peak of SN 1998bw is $B-V = 0.47 \pm 0.07$ mag (Galama et al. 1998a) while
the usual value for Type Ia events is $0.00 mag$ (Hamuy et al. 1996b). So
in all, the light curve of SN 1998bw looks similar to that of Type Ia
events, yet detailed parameters are quantitatively different.  

Perhaps a closer match can be found with the anomalous Type Ia SN 1991bg
(Leibundgut et al. 1993; Fillipenko et al. 1992). This event had a
substantially redder peak color ($B-V \sim 0.8$), a much lower peak 
absolute magnitude ($M_{B} = -16.62$), and no bump in the I band light
curves. While the detailed light curve (and spectrum) of SN 1991bg is
still different from that SN 1998bw, we note that many of the properties 
are more like those of SN 1991bg than of normal Type Ia events. 

Type Ib and Ic light curves have not been characterized as closely as
those of Type Ia supernovae. Nevertheless, enough is known (e.g., Uomoto
\& Kirshner, 1986; Ensman and Woosley 1988; Clocchiatti et al. 1997) to
find similarities and differences with SN 1998bw. The overall light 
curve of Type Ib and Ic events is the same for SN 1998bw with similar
decline rates over the first 15 days. The late time decline rate of 
Type Ic events vary, apparently with two classes, as slow and fast
decliners. The 60-180
day decline rates of roughly 0.016 magnitudes per day are seen for
the Type Ic events SN1983N and SN1983V (Clocchiatti et al. 1997) which is
comparable to that for SN 1998bw. The color evolution for Type Ic events
is similar to that of SN 1998bw, both at peak and at late times.
The peak absolute magnitude of Type Ic events vary about $M_{B} \sim
-17.5$, yet are all significantly fainter than SN 1998bw. However,
with the few well measured Type Ic events having a wide scatter, the
luminosity of SN 1998bw may not be unusual. In all, SN 1998bw appears 
to have a light curve within the class of Type Ic events.

Type II supernovae vary greatly in their light curve shape and color
(Patat et al. 1993). The colors, peak absolute magnitudes, decline rate
of SN 1998bw from peak, and late time decline rate are all within the
normal range for the IIL subclass. Nevertheless, there are some subtle
distinctions; such as a total lack of any indication of a plateau in the I
band and the switch to the late time decline rate only $\sim 30$ days
after peak. 

In all, the light curve of SN1998bw is fully consistent with those
of Type Ic supernovae, in keeping with the spectral classification and
physical models.

\section{Comparison with Models}

The decay rate of the tail is so close to an exponential that we suggest 
that this is no coincidence. In addition, the measured decline rate
corresponds to that expected from the decay of radio active cobalt (with a
half-life of 78.5 days) as modified by the effects due to the expansion of 
the shell (Colgate and McKee 1969). Hence, it is reasonable to take our
light curve as strong evidence that the late time light curve of SN 1998bw
is being powered by the decay of cobalt, with the difference in slope
caused by the leakage of gamma radiation from the shell. 

 Three detailed models have been presented seeking to explain the
light curve of SN1998bw. Iwamoto et al. (1998) and Iwamoto (1999) model
the event as an extremely energetic explosion of a massive star stripped
down to its carbon/oxygen core.  Woosley, Eastman, and Schmidt (1998)
independently present a similar model with similar results.
H\"{o}flich, Wheeler, and Wang (1998) and Wheeler, H\"{o}flich, and Wang
(1999) present a model with an aspherical explosion in the nondegenerate
C/O core of a massive star.  An asymmetric event can account for the
observed polarization.  All three models account for the early light
curve, the early colors, and the early spectrum with generally 
acceptable accuracy.

       Iwamoto et al. (1998) and Iwamoto (1999) present
predictions for the late-time V light curve of SN1998bw (see Figure 1 of
Iwamoto 1999) as a perfectly straight line in a log-log plot of flux
versus time since the burst, for a predicted power law with slope -2.75.
This power law prediction does not agree with the observed exponential
decline.  However, K. Nomoto (1999, private communication) has presented a
more detailed light curve prediction which shows a more complicated shape
(neither a power law nor an exponential) than presented in Iwamoto (1999).
The model V magnitude declines by 2.80 magnitudes from 27 June to 28
October with deviations from a simple exponential curve defined by the end
points of up to 0.22 mag.  For comparison, our data shows a decline of
2.26 mag over this same time and maximum departures from a simple
exponential decline of $<0.05 mag$.  

One way to distinguish the three models is by the explosion energy
and the ejected $^{56}Ni$ mass.  The spherically symmetric models have
energies and nickel masses of around $3 \times 10^{52} erg$ and $0.7
M_{\odot}$, while the aspherical models have $2 \times 10^{51} erg$ and
$0.2 M_{\odot}$.  Wheeler, H\"{o}flich, and Wang point out
that if the late time light curve tracks the radioactive decay line then
the ejected nickel mass can be determined, and that this might prove the
simplest discriminant between models.

        One possible method to measure the nickel mass from our light
curve is to scale the luminosity in the tail from another supernova of
known late-time luminosity and nickel mass.  The best case for comparison
might be SN1987A which has a well measured light curve (Hamuy et al.
1988), a well known distance (50 kpc; McCray 1993), and a well known
nickel mass ($0.069 \pm 0.003 M_{\odot}$; McCray 1993).  On day 170 after
the core
collapse, SN1998bw had an extinction corrected V magnitude of 17.60, while
SN1987A had an extinction corrected magnitude of 4.39.  If SN1987A were
placed at 39 Mpc, then its V magnitude should appear 1.25 mag fainter than
we observed for SN1998bw.  This implies a nickel mass 3.2 times larger, or
that SN1998bw has $0.22 M_{\odot}$.  The dominant uncertainty arises from
the distance to SN1998bw, for which peculiar velocities of up to $400 km
\cdot s^{-1}$ and uncertainties in the Hubble Constant of up to $10 km
\cdot s^{-1} \cdot Mpc^{-1}$ yield
nickel mass uncertainties of $0.09 M_{\odot}$.

        The procedure in the previous paragraph can only be approximate,
in particular since there might be substantial leakage of the gamma rays
from the expanding nebula.  Such leakage could explain why our observed
decline is steeper than that associated with $^{56}Co$ decay (Clocchiatti
\& Wheeler 1997). Such leakage would lower the late-time luminosity 
and lower the deduced nickel mass. The V light curve of SN1987A declined 
with the $^{56}Co$ rate whereas
the V light curve of SN1998bw declined at a roughly twice the rate.  The
effect of leakage on our previous derived nickel mass can only be
estimated within specific models, yet it is likely that our $0.22 \pm 0.09
M_{\odot}$ value must be regarded as a lower limit.

        We expect that our late-time light curve will provide a set of
observations useful for refining and constraining individual models of the
unique SN1998bw.  In particular, our data might constrain the quantity of
ejected $^{56}Ni$ so as to decide between symmetric and asymmetric models.
Another challenge to models is to explain the near perfect exponential
shape of the light curve  even though the slope is not that of the
$^{56}Co$ decay.

        We will continue to monitor the brightness of SN1998bw in 1999.
However, the background light from its host galaxy is increasingly a
problem for exact photometry.

\begin{figure}
\begin{center}
\resizebox{15cm}{14cm}{\includegraphics{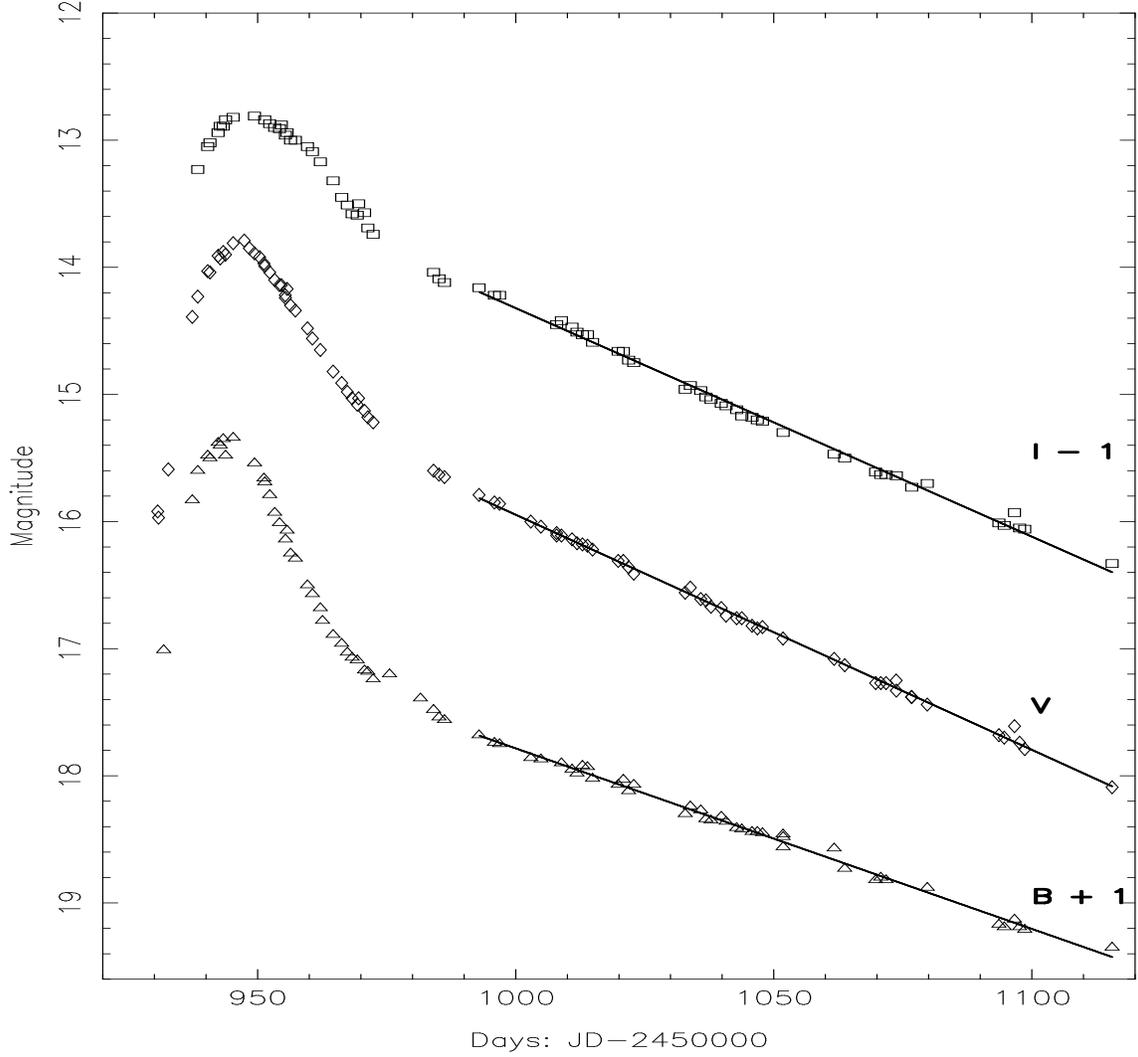}}
\end{center}
\caption{B, V, and I light curves for SN 1988bw. Galama et al. (1998) data
are all from before day 986 (JD2450986) while ours begin on day 992
(JD2450992). The best fit lines show a remarkably good fit to our measured
late time observations with no significant deviations at any time or in 
any color. The corresponding decay half-lives are $53.4 \pm 0.8$, $40.9
\pm 0.7$, and $41.6 \pm 0.7$ days in B, V, and I respectively. The
closeness of the magnitude-versus-time relation to a line and the
similarity of the decline rates with those for Type Ia supernovae are
suggestive that the decay of radioactive cobalt might be the source 
powering the tail with leakage of gamma radiation. The overall light curve 
shape is similar to those of
Type Ia, Ib, Ic and IIL supernovae. Our measurements with greater than 0.1
mag are not represented in this chart (see Table 1).}
\end{figure}

\begin{table}
\begin{center}
{\bf Table 1: SN 1998bw Light Curve.}
\\
\begin{tabular}{|cccc|}   
\hline
JD-2450000 & B & V & I \\
\hline
992.89 & $16.68 \pm 0.03$ & $15.79 \pm 0.02$ & $15.16 \pm 0.03$ \\
995.88 & $16.74 \pm 0.03$ & $15.85 \pm 0.02$ & $15.22 \pm 0.02$ \\
996.88 & $16.75 \pm 0.03$ & $15.86 \pm 0.02$ & $15.22 \pm 0.02$ \\
1002.92 & $16.86 \pm 0.03$ & $16.00 \pm 0.03$ & $\cdots$ \\
1004.88 & $16.87 \pm 0.03$ & $16.04 \pm 0.03$ & $\cdots$ \\
1007.93 & $\cdots$ & $16.09 \pm 0.03$ & $15.45 \pm 0.03$ \\
1007.94 & $\cdots$ & $16.11 \pm 0.03$ & $\cdots$ \\
1008.87 & $16.90 \pm 0.04$ & $16.11 \pm 0.03$ & $15.42 \pm 0.04$ \\ 
1010.90 & $16.95 \pm 0.03$ & $16.14 \pm 0.03$ & $15.47 \pm 0.03$ \\
1011.88 & $16.98 \pm 0.04$ & $16.17 \pm 0.02$ & $15.51 \pm 0.03$ \\
1012.92 & $16.92 \pm 0.03$ & $16.18 \pm 0.03$ & $15.53 \pm 0.03$ \\
1013.90 & $16.93 \pm 0.03$ & $16.19 \pm 0.03$ & $15.53 \pm 0.03$ \\
1014.90 & $17.02 \pm 0.03$ & $16.22 \pm 0.02$ & $15.59 \pm 0.03$ \\
1019.87 & $17.07 \pm 0.03$ & $16.31 \pm 0.02$ & $15.66 \pm 0.03$ \\
1020.86 & $17.03 \pm 0.03$ & $16.31 \pm 0.03$ & $15.66 \pm 0.03$ \\
1021.86 & $17.12 \pm 0.03$ & $16.36 \pm 0.02$ & $15.73 \pm 0.03$ \\
1022.86 & $17.07 \pm 0.03$ & $16.41 \pm 0.03$ & $15.75 \pm 0.03$ \\
1032.84 & $17.30 \pm 0.05$ & $16.56 \pm 0.03$ & $15.96 \pm 0.03$ \\
1033.84 & $17.24 \pm 0.07$ & $16.52 \pm 0.04$ & $15.93 \pm 0.04$ \\
1035.84 & $17.27 \pm 0.04$ & $16.61 \pm 0.03$ & $15.97 \pm 0.03$ \\
1036.84 & $17.34 \pm 0.03$ & $16.62 \pm 0.03$ & $16.02 \pm 0.03$ \\
1037.83 & $17.35 \pm 0.03$ & $16.67 \pm 0.02$ & $16.04 \pm 0.03$ \\
1039.82 & $17.32 \pm 0.03$ & $16.68 \pm 0.02$ & $16.07 \pm 0.03$ \\
1040.77 & $17.36 \pm 0.03$ & $16.74 \pm 0.02$ & $16.09 \pm 0.03$ \\
1041.80 & $\cdots$ & $16.75 \pm 0.02$ & $\cdots$ \\
1041.84 & $17.39 \pm 0.03$ & $16.84 \pm 0.04$ & $\cdots$ \\
\hline
\end{tabular}
\end{center}
\end{table}

\begin{table}   
\begin{center}  
\begin{tabular}{|cccc|}
\hline
1042.82 & $17.41 \pm 0.03$ & $16.76 \pm 0.02$ & $16.12 \pm 0.03$ \\
1043.81 & $17.42 \pm 0.03$ & $16.76 \pm 0.02$ & $16.17 \pm 0.03$ \\
1045.80 & $17.44 \pm 0.03$ & $16.82 \pm 0.02$ & $16.18 \pm 0.03$ \\
1046.79 & $17.43 \pm 0.03$ & $\cdots$ & $\cdots$ \\
1046.80 & $17.44 \pm 0.03$ & $16.84 \pm 0.02$ & $16.20 \pm 0.03$ \\
1047.81 & $17.45 \pm 0.03$ & $16.83 \pm 0.02$ & $16.21 \pm 0.03$ \\
1051.79 & $17.46 \pm 0.03$ & $16.92 \pm 0.02$ & $\cdots$ \\
1051.80 & $17.48 \pm 0.03$ & $\cdots$ & $\cdots$ \\
1051.80 & $17.56 \pm 0.03$ & $\cdots$ & $16.30 \pm 0.03$ \\
1061.70 & $17.57 \pm 0.06$ & $17.08 \pm 0.04$ & $16.47 \pm 0.03$ \\
1062.76 & $17.57 \pm 0.15$ & $16.85 \pm 0.15$ & $\cdots$ \\
1062.76 & $\cdots$ & $16.92 \pm 0.15$ & $\cdots$ \\
1063.75 & $17.73 \pm 0.04$ & $17.13 \pm 0.03$ & $16.50 \pm 0.03$ \\ 
1064.76 & $17.61 \pm 0.13$ & $17.28 \pm 0.18$ & $16.75 \pm 0.27$ \\
1069.73 & $17.82 \pm 0.03$ & $17.27 \pm 0.02$ & $16.61 \pm 0.03$ \\
1070.73 & $17.80 \pm 0.03$ & $17.27 \pm 0.02$ & $16.63 \pm 0.03$ \\
1071.76 & $17.82 \pm 0.04$ & $17.27 \pm 0.03$ & $16.63 \pm 0.03$ \\
1073.72 & $\cdots$ & $17.25 \pm 0.03$ & $\cdots$ \\
1073.73 & $\cdots$ & $17.33 \pm 0.03$ & $16.64 \pm 0.03$ \\
1076.70 & $\cdots$ & $17.38 \pm 0.03$ & $\cdots$ \\
1076.71 & $\cdots$ & $17.38 \pm 0.03$ & $16.73 \pm 0.03$ \\
1079.73 & $17.88 \pm 0.03$ & $17.44 \pm 0.03$ & $16.70 \pm 0.03$ \\
1093.62 & $18.17 \pm 0.04$ & $17.68 \pm 0.03$ & $17.01 \pm 0.03$ \\
1094.63 & $18.19 \pm 0.04$ & $17.70 \pm 0.03$ & $17.03 \pm 0.03$ \\
1096.63 & $18.13 \pm 0.05$ & $17.61 \pm 0.04$ & $16.93 \pm 0.05$ \\
1097.63 & $18.19 \pm 0.03$ & $17.74 \pm 0.03$ & $17.05 \pm 0.03$ \\
1098.62 & $18.21 \pm 0.03$ & $17.79 \pm 0.03$ & $17.06 \pm 0.03$ \\
1115.55 & $18.35 \pm 0.06$ & $18.09 \pm 0.05$ & $17.33 \pm 0.06$ \\
\hline
\end{tabular}
\end{center}
\end{table}

\end{document}